\documentclass[aps, prd, superscriptaddress, twocolumn, 10pt]{revtex4-1}

\usepackage{amsmath}	
\usepackage{amsthm}		
\usepackage{amssymb}	
\usepackage{array}
\usepackage{datetime}
\usepackage{graphicx}
\usepackage{color}
\usepackage{verbatim}
\usepackage{bm}
\usepackage{braket}
\usepackage{natbib}
\usepackage{cancel}
\usepackage{xcolor}
\usepackage{slashed}
\usepackage{braket}
\usepackage{amsfonts} 
\smallskipamount
\usepackage{feynmp}
\usepackage{letltxmacro} 
\usepackage[colorlinks=true, citecolor=midblue, linkcolor=midblue, urlcolor=midblue]{hyperref}

\usepackage{setspace}
\usepackage[normalem]{ulem}

\usepackage{tikz,xcolor}
\definecolor{lime}{HTML}{A6CE39}
\DeclareRobustCommand{\orcidicon}{
	\begin{tikzpicture}
	\draw[lime, fill=lime] (0,0) 
	circle [radius=0.16] 
	node[white] {{\fontfamily{qag}\selectfont \tiny ID}};
	\draw[white, fill=white] (-0.0625,0.095) 
	circle [radius=0.007];
	\end{tikzpicture}
	\hspace{-2mm}
}
\foreach \x in {A, ..., Z}{
	\expandafter\xdef\csname orcid\x\endcsname{\noexpand\href{https://orcid.org/\csname orcidauthor\x\endcsname}{\noexpand\orcidicon}}
}

\settimeformat{ampmtime}

\definecolor{grey}{rgb}{0.4,0.4,0.4}
\definecolor{dullmagenta}{rgb}{0.4,0,0.4}
\definecolor{darkblue}{rgb}{0,0,0.4}
\definecolor{midblue}{rgb}{0,0,0.5}
\definecolor{midred}{rgb}{0.5,0,0}
\definecolor{orange}{rgb}{1,0.5,0}
\definecolor{lightbrown}{rgb}{0.75,0.5,0.25}
\definecolor{tan}{cmyk}{0.14,0.42,0.56,0}
\definecolor{djunglegreen}{cmyk}{0.99,0,0.52,0}
\definecolor{lightgreen}{rgb}{0,1,0}
\definecolor{olivegreen}{cmyk}{0.64,0,0.95,0.40}
\definecolor{midgreen}{rgb}{0.0,0.675,0.0}
\definecolor{darkgreen}{rgb}{0,0.5,0}

\newcommand{\vs}{\vspace}

\renewcommand{\.}{\hspace{0.5mm}}

\newcommand{\Erm}{\ensuremath{\mathrm{E}}}

\newcommand{\Hrm}{\ensuremath{\mathrm{H}}}

\newcommand{\Prm}{\ensuremath{\mathrm{P}}}

\newcommand{\Srm}{\ensuremath{\mathrm{S}}}

\newcommand{\frm}{\ensuremath{\mathrm{f}}}
\newcommand{\grm}{\ensuremath{\mathrm{g}}}

\newcommand{\srm}{\ensuremath{\mathrm{s}}}

\newcommand{\Ocal}{\ensuremath{\mathcal{O}}}

\renewcommand{\d}{\ensuremath{\mathrm{d}}}

\settimeformat{ampmtime}

\makeatletter
\let\oldr@@t\r@@t
\def\r@@t#1#2{%
\setbox0=\hbox{$\oldr@@t#1{#2\,}$}\dimen0=\ht0
\advance\dimen0-0.2\ht0
\setbox2=\hbox{\vrule height\ht0 depth -\dimen0}%
{\box0\lower0.4pt\box2}}
\LetLtxMacro{\oldsqrt}{\sqrt}
\renewcommand*{\sqrt}[2][\ ]{\oldsqrt[#1]{#2}}
\makeatother

\newcommand{\FirstAffiliation}{\affiliation{
	Arnold Sommerfeld Center,
	Ludwig-Maximilians-Universit{\"a}t,
	Theresienstra{\ss}e 37,
	80333 M{\"u}nchen,
	Germany}}

\newcommand{\ThirdAffiliation}{\affiliation{
	Fakult{\"a}t Physik,
	Technische Universit{\"a}t Dortmund,
	August-Schmidt-Str.~4,
	44221 Dortmund,
	Germany}}

\newcommand{\FourthAffiliation}{\affiliation{Universit{\'e} Paris--Saclay, CNRS, CEA, Institut de Physique Theorique, 91191, Gif-sur-Yvette, France
}}

\begin{document}
	
\title{Observing Micro Black Hole Dark Matter}

\author{Manuel Ettengruber\!\orcidE}
\email{manuel-meinrad.ettengruber@ipht.fr}
\FourthAffiliation

\author{Florian K{\"u}hnel\!\orcidC}
\email{fkuehnel@mpp.mpg.de}
\FirstAffiliation
\ThirdAffiliation

\date{\formatdate{\day }{ \month }{ \year}, \currenttime}

\begin{abstract}
Primordial micro black holes can constitute dark matter if short-distance gravity is modified by extra dimensions or a large number of species and if the memory-burden effect sufficiently suppresses Hawking evaporation. The resulting black holes in the transition regime differ from their four-dimensional Einsteinian counterparts through their mass--radius relation, temperature, entropy, and lifetime, which can render even very light objects cosmologically stable. The most promising observational consequences of such micro black hole dark matter are analysed. Neutron star survival yields the most robust constraints, while a narrow region of parameter space can simultaneously remain viable and address the missing-pulsar problem in the Galactic center. Diffuse evaporation signals in neutrino telescopes are found to be relevant mainly in extra-dimensional scenarios, whereas in generic species models, visible emission is strongly suppressed by evaporation into dark sectors. Merger-induced evaporation bursts can provide an additional probe in extra-dimensional realisations if the post-merger remnant briefly returns to the semiclassical phase. Overall, micro black hole dark matter remains phenomenologically viable in constrained regions, with neutron stars, neutrino telescopes, and merger signatures providing complementary tests.
\end{abstract}

\maketitle

\section{Introduction}
\label{sec:Introduction}

\noindent Identifying the origin of dark matter (DM) remains one of the central open problems in fundamental physics. While a broad range of particle and compact-object candidates has been proposed, the lack of decisive non-gravitational evidence keeps the space of viable models wide. Primordial black holes (PBHs)~\cite{Zeldovich:1967lct, Carr:1974nx, Carr:1975qj} (see Refs.~\cite{Carr:2016drx, Carr:2020xqk, Escriva:2022duf} for reviews) are among the most intriguing possibilities, since they {\it a priori} require no extension of the Standard Model particle content in order to account for the dark matter abundance. In their conventional four-dimensional Einsteinian realisation, however, PBH dark matter is subject to strong observational constraints over a wide range of masses; in particular, sufficiently light black holes are expected to evaporate on timescales much shorter than the age of the Universe.

A qualitatively different situation can arise if gravity is modified at short distances. One important indication in this direction is the species scale of gravity \cite{Dvali:2007hz, Dvali:2007wp}, according to which the fundamental scale at which gravitational interactions become strongly coupled is lowered in the presence of many particle species,
\begin{equation}
    M_{\frm}
         = 
            \frac{ M_{\Prm} }{ \sqrt{N} }
            \, ,
            \label{eq:masterequation}
\end{equation}
where $M_{\Prm}$ is the Planck scale and $N$ denotes the number of species. This relation has far-reaching implications. It implies that the onset of genuinely quantum-gravitational physics can occur well below the Planck scale, even in otherwise weakly coupled low-energy theories. The best-known realisation of this idea is the scenario of large extra dimensions~\cite{Arkani-Hamed:1998jmv, Arkani-Hamed:1998sfv, Antoniadis:1998ig}, in which the large multiplicity of Kaluza--Klein states lowers the fundamental gravitational scale, potentially even to the TeV range. More general species frameworks extend the same logic to theories with large dark sectors or other hidden particle multiplicities~\cite{Dvali:2009ne, Arkani-Hamed:2016rle, Arkani-Hamed:2020yna}.

Such theories are accompanied by a broad phenomenology. Small neutrino masses can arise naturally in extra-dimensional and species settings through mixing with a large number of additional states~\cite{Arkani-Hamed:1998wuz, Dvali:2009ne, Ettengruber:2022pxf, Ettengruber:2025usk}. The same underlying structure can lead to distinctive signatures in neutrino oscillation experiments~\cite{Dvali:1999cn, Ettengruber:2022pxf, Ettengruber:2025usk, Machado:2011jt, Machado:2011kt, Basto-Gonzalez:2012nel, Girardi:2014gna, Rodejohann:2014eka, Berryman:2016szd, Carena:2017qhd, Stenico:2018jpl, Arguelles:2019xgp, DUNE:2020fgq, Basto-Gonzalez:2021aus, Arguelles:2022xxa, Ettengruber:2024fcq, Eller:2025lsh, Antoniadis:2025rck, Elacmaz:2025ihm, Bai:2026kdq}, and in species theories analogous effects may also arise in other sectors such as neutron physics~\cite{Dvali:2009ne, Dvali:2023zww}. Collider probes and astrophysical tests have likewise been discussed extensively~\cite{Arkani-Hamed:1998sfv, Giudice:1998ck, Hall:1999mk, Abazajian:2000hw, Hanhart:2001fx, Hannestad:2003yd, Dvali:2009fw, Cohen:2018cnq, DAgnolo:2019cio,ATLAS:2021kxv, CMS:2021far,Zander:2023jcu, Ettengruber:2023tac, Ettengruber:2025kat, DAgnolo:2025cxb}. Among all such probes, however, black holes occupy a particularly privileged position, since they are the most direct manifestation of strong gravitational dynamics and their properties are sensitive to gravity in a largely model-independent way.

In theories with a lowered fundamental scale, black holes are expected to exhibit a transition regime between the fully quantum domain at distances of order
\begin{equation}
    r_{\frm}
         = 
            \frac{ 1 }{ M_{\frm} }
\end{equation}
and the ordinary four-dimensional Einsteinian regime. Denoting by $R$ the length scale at which the transition to standard gravity is completed, one expects~\cite{Dvali:2008fd, Dvali:2008rm}
\begin{equation}
    r_{\frm}
        \ll
            r
        \ll
            R
            \, .
    \label{eq:lengthscalerange}
\end{equation}
In extra-dimensional models, $R$ admits the geometric interpretation of the compactification radius and is constrained by short-distance tests of Newtonian gravity, currently at the level of $R \lesssim 30\,\mu{\rm m}$~\cite{Lee:2020zjt}. In terms of black hole masses, the transition regime corresponds to
\begin{equation}
    M_{\frm}
        \ll
    M
        \ll
            \begin{cases}
                \sqrt{N}\,M_{\Prm}
                \qquad & (\text{species})
                \\[2mm]
                ( M_{\Prm}\.R )\,M_{\Prm}
                \qquad & (\text{extra dimensions})
            \end{cases}
            \, .
            \label{eq:Micro-Black-Hole-Regime}
\end{equation}
Black holes in this range differ from their ordinary four-dimensional counterparts in their mass--radius relation, temperature, entropy, and lifetime. As a result, very light black holes can have properties that are qualitatively different from those of standard evaporating Schwarzschild black holes.

This possibility becomes particularly interesting in view of recent work on the memory-burden effect~\cite{Dvali:2018xpy, Dvali:2020wft, Alexandre:2024nuo, Dvali:2024hsb, Thoss:2024hsr, Boccia:2025hpm, Chianese:2024rsn, Chianese:2025wrk, Calabrese:2025sfh, Montefalcone:2025akm, Dvali:2025ktz}. The core of this effect is that systems of high information capacity undergo a universal back reaction. In this picture, Hawking evaporation can become strongly suppressed after the Page time, leading to a substantial enhancement of black hole lifetimes. Since the suppression is governed by powers of the entropy, its impact is modified precisely in the short-distance transition regime where the entropy itself deviates from the four-dimensional Einsteinian expression. This interplay has recently been studied both in extra-dimensional realisations~\cite{Anchordoqui:2024dxu} and, more generally, for theories with modified short-distance gravity~\cite{Ettengruber:2025kzw}. A striking consequence is that black holes with masses far below the ordinary Planck scale, potentially as low as $10^{14}\,{\rm GeV}$, can remain cosmologically stable. In the following we refer to such objects as micro black holes, or $\mu$BHs.

If $\mu$BHs are stable on cosmological timescales and make up the dark matter, their phenomenology differs radically from that of conventional PBH scenarios. Their tiny masses imply enormous number densities, so that encounters with astrophysical objects, terrestrial detectors, and other black holes become frequent enough to be potentially observable. For illustration, if the dark matter consisted of black holes with masses of order $10^{14}\,{\rm GeV}$, their flux through ordinary macroscopic objects would be very large. This raises an immediate question: is such a scenario already excluded, or can it survive existing observational constraints while giving rise to distinctive signals?

In this work we address this question systematically. We study primordial micro black holes as dark matter in theories with modified short-distance gravity, focusing on two broad classes of realisations: extra-dimensional models and generic species theories. We assume throughout that the memory-burden effect governs the late-time evaporation of the black holes and investigate the resulting observational consequences. Our main goal is to identify the leading constraints and the most promising discovery channels for $\mu$BH dark matter.

We find that neutron star survival provides the strongest and most robust probe of the scenario. At the same time, a narrow region of parameter space can remain viable while potentially addressing the missing-pulsar problem in the Galactic center~\cite{2010ApJ...715..939M, 2012ApJ...753..108W, Dexter:2013xga, 2014ApJ...780L...3S}. Diffuse evaporation signals in neutrino telescopes are relevant mainly in extra-dimensional realisations, whereas in generic species models the visible emission is strongly suppressed by evaporation into dark sectors. We also consider merger-induced evaporation bursts, which can arise if the remnant of a merger briefly returns to the semiclassical phase before re-entering the memory-burden regime. Purely gravitational signatures are briefly discussed for completeness, although they appear less promising with present or near-future sensitivities.

The paper is organised as follows. In Sec.~\ref{sec:Setup} we review the relevant properties of black holes in the transition regime and summarise the ingredients of the memory-burden framework. In Sec.~\ref{sec:Evaporating-Micro-black holes} we analyse the direct evaporation signal from an ambient population of $\mu$BHs. In Sec.~\ref{sec:Neutron star-Consumption-by-Micro-black holes} we derive constraints from neutron star survival. In Sec.~\ref{sec:Other-Gravitational-Effects} we comment on additional gravitational signatures. In Sec.~\ref{sec:Micro-black holes-Mergers} we study merger-induced signals and estimate the resulting flux of evaporation products at Earth. We conclude in Sec.~\ref{sec:Conclusion}.

\section{Setup}
\label{sec:Setup}

\noindent In this section we summarise the properties of black holes in the intermediate regime~\eqref{eq:Micro-Black-Hole-Regime}, where gravity departs from its ordinary four-dimensional Einsteinian behavior but remains outside the fully quantum domain. The central modification in this regime is the mass--radius relation. Instead of the standard four-dimensional Schwarzschild scaling, the black hole radius takes the form~\cite{Argyres:1998qn, Dvali:2008fd}
\begin{equation}
    r_{\Srm}^{({\tilde n})}
         = 
            \frac{ a_{\tilde n} }{ M_{\frm} }\mspace{-2mu}
            \left(
                \frac{ M }{ M_{\frm} }
            \right)^{\!1/(1 + {\tilde n})}
            ,
    \label{eq:r-S}
\end{equation}
where ${\tilde n}$ parametrises the short-distance deformation of gravity and $a_{\tilde n}$ is an $\Ocal( 1 )$ coefficient. In extra-dimensional models, ${\tilde n}$ is identified with the number of extra dimensions, and in that case we denote it by $n$ throughout. In more general species frameworks, ${\tilde n}$ is determined by the effective geometry of the species space ~\cite{Dvali:2008fd}. For extra-dimensional models the $a_{\tilde n}$ is given by 
\begin{equation}
    a_{n}
        =
            \left(
                8\mspace{1mu}\pi^{-(n+1)/2}\,
                \frac{
                \Gamma
                \big[
                    (n+3)/2\mspace{1mu}
                \big]}{n+2}
            \right)^{\!1/(1+n)}
            \, .
            \label{eq:a-n}
\end{equation}
In species theories the parameter $a_{\tilde{n}}$ in Eq.~\eqref{eq:r-S} is unity~\cite{Dvali:2008fd}. 

Once the radius is modified, all thermodynamic properties of the black hole are correspondingly altered. In the extra-dimensional case, the Hawking temperature becomes~\cite{Myers:1986un, Myers:1986rx, Argyres:1998qn, Friedlander:2022ttk}
\begin{equation}
    T_{\Hrm}
         = 
            \frac{ n + 1 }{ 4\mspace{1mu}\pi\,r_{\Srm}^{(n)} }
            \, ,
            \label{eq:T-H}
\end{equation}
which implies the evaporation rate
\begin{subequations}
\begin{equation}
    \frac{ \d M }{ \d t }\bigg|_{\rm extra}
        = 
            -\mspace{2mu}\alpha\bigl( n,\mspace{1.5mu}T_{\Hrm} \bigr)\,
            T_{\Hrm}^{2}
            \, .
            \label{eq:dMdt-Extra-Dimensions}
\end{equation}
Here $\alpha$ encodes the effective number of thermally accessible degrees of freedom. Its precise form in the higher-dimensional regime can be found in Ref.~\cite{Cardoso:2005mh, Friedlander:2022ttk}. In the parameter range relevant for our discussion, the temperature is sufficiently high that $\alpha$ can be treated, to a good approximation, as constant.

In species theories, the enhanced number of accessible states modifies the evaporation law differently and yields~\cite{Dvali:2008fd}
\begin{equation}
    \frac{\d M}{\d t}\bigg|_{\rm species}
         = 
            -\mspace{2mu}M_{\frm}^{2}
            \left(
                \frac{ M_{\frm} }{ M }
            \right)^{\!(2 - {\tilde n})/(1 + {\tilde n})}
            \, .
            \label{eq:dMdt-Species}
\end{equation}
\end{subequations}

This expression includes evaporation into all available species, including dark-sector states. If one is interested only in the fraction of the emitted energy deposited into the visible sector, the result must be rescaled by~\cite{Dvali:2008fd}
\begin{equation}
    \frac{ E_{{\rm BH}\,\longrightarrow\,\text{our sector}} }
    { E_{{\rm BH}\,\longrightarrow\,\text{all sectors}} }
        \sim
            \left(
                \frac{ M_{\frm} }{ M }
            \right)^{\!{\tilde n}/(1 + {\tilde n})}
            \, .
            \label{eq:E-Ratio}
\end{equation}
The root of the above equations lies in the observation that, depending on its
horizon size in species space, a black hole evaporates into all species accessible
to it. As the black hole becomes heavier, it is able to evaporate into more and
more species, up to the maximum number $N$, which is of course the total number
of fundamental species. The equations above are derived for the case of $N$ copies
of species related by an exact permutation symmetry~\cite{Dvali:2008fd}.

Integrating the semiclassical evaporation law gives the corresponding lifetimes. In extra dimensions one obtains~\cite{Argyres:1998qn}
\begin{subequations}
\begin{equation}
    \tau_{\rm extra}\Big|_{\rm SC}
         = 
            \frac{ C_{\rm ADD} }{ M_{\frm} }\mspace{-2mu}
            \left(
                \frac{ M }{ M_{\frm} }
            \right)^{\!(n + 3)/(n + 1)}
            ,
            \label{eq:tau-Extra-Dimensions}
\end{equation}
with $C_{\rm ADD} \equiv 16\mspace{1mu}\pi^{2}\.a_{n}^{2} / [ \alpha\.(n + 1)(n + 3) ]$, whereas in species theories one finds~\cite{Dvali:2008fd}
\begin{equation}
    \tau_{\rm species}\Big|_{\rm SC}
         = 
            \frac{ C^{(N)} }{ M_{\frm} }\mspace{-2mu}
            \left(
                \frac{ M }{ M_{\frm} }
            \right)^{\!3/({\tilde n} + 1)}
            ,
            \label{eq:tau-Species}
\end{equation}
\end{subequations}
with $C^{(N)} \equiv ({\tilde n} + 1) / 3$.

A second key quantity is the entropy. Since the entropy is determined by the horizon area, the modified radius implies a modified entropy as well. In the extra-dimensional case this gives~\cite{Argyres:1998qn}
\begin{subequations}
\begin{align}
    S_{\rm extra}
        & = 
            S^{(4)}
            \left(
                \frac{ R }{ r_{\Srm} }
            \right)^{\!n/(n + 1)}
            \notag
            \\[2mm]
        &\sim
            S^{(4)}
            \left(
                \frac{ M_{\Prm}^{2n + 2} }
                { M_{\frm}^{n + 2}\.M^{n} }
            \right)^{\!1/(n + 1)}
            ,
            \label{eq:S-Extra-Dimensions}
\end{align}
\end{subequations}
where $r_{\Srm}$ and $S^{(4)}$ denote the ordinary four-dimensional Schwarzschild radius and entropy, respectively. In the species case, the corresponding scaling is~\cite{Ettengruber:2025kzw}
\begin{equation}
    S_{\rm species}
        \sim
            S^{(4)}
            \left(
                \frac{ M_{\Prm} }{ M }
            \right)^{\!2{\tilde n}/({\tilde n} + 1)}
            N^{\!1/({\tilde n} + 1)}
            \, .
            \label{eq:S-Species}
\end{equation}

In the context of higher-dimensional theories, the phenomenology of the associated black hole properties has been studied over the past two decades (see e.g.~Refs.~\cite{Conley:2006jg, Friedlander:2022ttk, Anchordoqui:2022txe, Anchordoqui:2024tdj}). Moreover, these entropy relations are even more important when we take the memory-burden effect into account as it suppresses evaporation by powers of the entropy. We parametrise the evaporation rate in the memory-burden phase as (cf.~Ref.~\cite{Dvali:2020wft})
\begin{equation}
    \frac{ \d M }{ \d t }\bigg|_{\rm MB}
         = 
            \frac{ 1 }{ S^{k} }\,
            \frac{ \d M }{ \d t }\bigg|_{\rm SC}
            \, ,
            \label{eq:dMdt-Memory-Burden}
\end{equation}
where $k$ characterises the strength of the suppression. In the scenarios of interest one typically expects~\cite{Dvali:2024hsb}
\begin{equation}
    1
        \leq
            k
        \leq
            3
            \, .
\end{equation}
Equation~\eqref{eq:dMdt-Memory-Burden} specifies the evaporation rate once the black hole has entered the memory-burden regime.

Besides the suppression exponent $k$, two additional quantities are relevant for the onset of the effect: the fraction $q$ of the black hole mass that must be lost before the memory burden becomes important, and the width $\delta$ of the transition into that regime. Theoretical arguments lead to~\cite{Dvali:2025ktz}
\begin{subequations}
\begin{align}
    q
        &\simeq
            \bigl(
                p^{2}\.S
            \bigr)^{-1/[2( p - 1 )]}
            \, ,
            \label{eq:q-S-p}
            \\[2mm]
    \delta
        & = 
            \frac{ q }{ ( p - 1 )\.\ln S }
        \simeq
            \frac{
            \bigl(
                p^{2}\.S
            \bigr)^{-1/[2( p - 1 )]} }
            { ( p - 1 )\.\ln S }
            \qquad
            (p \neq 1)
            \, .
            \label{eq:delta-S-p}
\end{align}
\end{subequations}
Here $p$ is the parameter controlling the microscopic structure of the memory-burden transition.

Using the entropy scalings above, one can translate these expressions into the extra-dimensional and species cases~\cite{Ettengruber:2025kzw},
\begin{subequations}
\begin{align}
    q_{\rm extra}
        &\propto
            \bigl(
                M_{\Prm}^{2}\,r_{\Srm}^{2}
            \bigr)^{-1/[2( p - 1 )]}
            \left(
                \frac{ r_{\Srm} }{ R }
            \right)^{\!n/[2( n + 1 )( p - 1 )]}
        ,
\end{align}
and
\begin{align}
    q_{\rm species}
        &\propto
            \left(
                \frac{ M_{\Prm} }{ M }
            \right)^{\!1/[( {\tilde n} + 1 )( p - 1 )]}
            N^{-1/[2( {\tilde n} + 1 )( p - 1 )]}
            \, ,
\end{align}
\end{subequations}
respectively. The final expressions for the lifetime of $\mu$BHs for the different cases read
\begin{subequations}
\begin{align}
    \tau_{\rm extra} 
        &\simeq 
            C_{\rm ADD}\.M_{\rm f}^{-(n+2)(k+2)/(n+1)} M^{[k(n+2)+n+3]/(n+1)}
        ,
        \label{eq:memorylifetimeextra}
\end{align}
and
\begin{align}
    \tau_{\rm species}
        &\simeq
           C^{(N)}\.M^{(3+2k)/(n+1)} M_\textrm{f}^{-(4+n+2k)/(n+1)}
            \, .
              \label{eq:memorylifetimespecies}
\end{align}
\end{subequations}

The relations collected above are the basic ingredients needed for the phenomenological analysis in the following sections. They determine the mass dependence of the radius, temperature, entropy, evaporation rate, and lifetime of micro black holes in the transition regime, both in the semiclassical phase and after the onset of memory burden. In particular, they make it possible for black holes with masses far below the ordinary four-dimensional evaporation threshold to survive on cosmological timescales and therefore to be viable dark matter candidates.

\section{Evaporating Micro Black Holes}
\label{sec:Evaporating-Micro-black holes}

\noindent A natural question is whether an ambient population of micro black holes can be detected through their evaporation products. The answer is not obvious a priori, because several competing effects are at work. On the one hand, if $\mu$BHs constitute the dark matter, their small masses imply very large number densities, and lighter black holes radiate more efficiently. On the other hand, the same objects must already be in the memory-burden phase in order to be cosmologically long-lived, which suppresses their evaporation rate. In addition, black holes in the transition regime are generally colder than ordinary four-dimensional Schwarzschild black holes of the same mass. The observable signal is therefore determined by a nontrivial interplay between abundance, temperature, and memory-burden suppression.

Neutrino telescopes provide a particularly interesting target for such a signal. If the typical energy of the emitted particles is of order the Hawking temperature, then the evaporation products can fall into an energy range accessible to experiments such as {\it IceCube} or {\it The Pacific Ocean Neutrino Experiment (P-ONE)}, depending on the fundamental scale and on the location in parameter space. Since individual $\mu$BHs are extremely light, the relevant observable is not a resolved long-lived source, but the cumulative signal produced by the ensemble of black holes present inside the detector volume at any given time.

We therefore estimate the characteristic waiting time for the emission of an observable particle from the population of $\mu$BHs inside a neutrino telescope. Assuming that all dark matter consists of micro black holes of mass $m_{\mu{\rm BH}}$, their number density is
\begin{equation}
    n^{}_{\mu{\rm BH}}
         = 
            \frac{ \rho_{\rm DM} }{ m^{}_{\mu{\rm BH}} }
            \, ,
\end{equation}
where $\rho_{\rm DM}$ denotes the local dark matter density. The corresponding flux through a detector of area $A$ is then
\begin{equation}
    \phi^{}_{\mu{\rm BH}}
         = 
            A\,v_{\rm DM}\,n^{}_{\mu{\rm BH}}
            \, ,
\end{equation}
with $v_{\rm DM}$ the local dark matter velocity.

To estimate the signal rate, we first determine the characteristic time over which a single black hole emits an amount of energy comparable to one Hawking quantum. In the memory-burden phase, the evaporation rate is given by \eqref{eq:dMdt-Memory-Burden} and using the scaling relations summarised in Sec.~\ref{sec:Setup} together with the results of Ref.~\cite{Ettengruber:2025kzw}, it yields for the extra-dimensional case
\begin{subequations}
\begin{align}
\begin{split}
    \frac{ \d M }{ \d t }\bigg|_{\rm MB}^{\rm extra}
        &= 
            -\mspace{2mu}\alpha_{\rm extra}
            \left(
                \frac{ n + 1 }{ 4\mspace{1mu}\pi }
            \right)^{\!2}
            \left(
                \frac{ M_{\frm} }{ a_{n} }
            \right)^{\!2}
            \\[2mm]
        &\phantom{ = \;}
            \times\mspace{-2mu}
            \left(
                \frac{ M_{\frm} }{ M }
            \right)^{\![k( n + 2 ) + 2]/( n + 1 )}
            ,
\end{split}
\end{align}
and
\begin{align}
    \frac{ \d M }{ \d t }\bigg|_{\rm MB}^{\rm species}
        & = 
            -\mspace{2mu}\alpha_{\rm species}
            \left(
               \frac{ M_{\frm} }{ M }
            \right)^{\!( 2k + 2 )/( n + 1 )}\mspace{-2mu}
            M_{\frm}^{2}
            \, .
\end{align}
\end{subequations}
for the species case, where we have written the rate relevant for observable Standard Model emission rather than the total evaporation rate into all sectors. This distinction is essential, because the large number of dark species strongly suppresses the visible signal even when the total emitted power is substantial.

Let $\tau_{\rm obs}$ denote the time required for a single $\mu$BH to emit an energy of order $T_{\Hrm}$. Estimating one observable quantum by the condition
\begin{equation}
    T_{\Hrm}
        = 
            \frac{ \d M }{ \d t }\bigg|_{\rm MB}
            \tau_{\rm obs}
            \, ,
\end{equation}
one obtains the corresponding detector waiting time by dividing by the number of black holes contained in the detector volume,
\begin{equation}
    \tau_{\rm det}
        = 
            \frac{ \tau_{\rm obs} }{ V\.n^{}_{\mu{\rm BH}} }
            \, ,
\end{equation}
where $V$ is the detector volume.

This leads to the following parametric estimate for the extra-dimensional case:
\begin{subequations}
\begin{equation}
    \tau_{\rm det}^{\rm extra}
        = 
            \frac{ a_{n} }{ \alpha_{\rm extra} }\.
            \frac{ 4\mspace{1mu}\pi }{ n + 1 }\.
            \frac{ 1 }{ V \rho_{\rm DM} }\.
            \frac{ M }{ M_{\star} }\mspace{-2mu}
            \left(
                \frac{ M }{ M_{\star} }
            \right)^{\![k( n + 2 ) + 1]/( n + 1 )}
            ,
\end{equation}
and, analogously, for the species case,
\begin{equation}
    \tau_{\rm det}^{\rm species}
        = 
            \frac{ 1 }{ \alpha_{\rm species}\,V \rho_{\rm DM} }\.
            \frac{ M }{ M_{\star} }\mspace{-2mu}
            \left(
                \frac{ M }{ M_{\star} }
            \right)^{\!(2k + 1)/( n + 1 )}
        .
\end{equation}
\end{subequations}

These expressions can be converted into an expected event rate for a detector with a given fiducial volume and exposure time. For the present purpose, however, the main issue is not only how often such events occur, but also how they would appear experimentally. A population of evaporating $\mu$BHs would generically produce a mixture of final states, many of which could easily be classified as background in existing searches. The cleanest channel is provided by charged evaporation products that generate Cherenkov light while traversing the detector volume. In contrast to ordinary astrophysical neutrino events, such signals would not be associated with a preferred arrival direction and would instead reflect the thermal energy scale set by the Hawking temperature of the underlying black hole population.

A realistic assessment of detector sensitivity would require a dedicated analysis of triggering, reconstruction, and background rejection, which lies beyond the scope of the present work. As a simple benchmark, we assume that a signal becomes potentially detectable if the expected event rate is at least of order one event per year over the operational lifetime of the detector. Under this assumption, the resulting sensitivity is shown in Fig.~\ref{fig:1} and Fig.~\ref{fig:2} for $k = 1$ and $k = 2$, respectively. The figures indicates that neutrino telescopes can probe a portion of the light-mass parameter space in extra-dimensional realisations, where the visible-sector emission remains appreciable and the Hawking temperature can lie in an experimentally-favourable range.

The situation is much less promising in generic species scenarios. There the total evaporation rate may still be sizable, but only a parametrically small fraction of the emitted power is deposited into Standard Model particles, since most of the radiation is lost into dark sectors. As a result, the diffuse evaporation signal in terrestrial detectors is strongly suppressed, and present neutrino telescopes are not expected to provide meaningful sensitivity in that case, as can observed in Fig.~\ref{fig:3}.

Direct searches for the evaporation products of ambient $\mu$BH dark matter are potentially relevant only in a restricted part of parameter space, primarily in extra-dimensional models. Even there, detectability is limited to the lightest viable black holes and depends on the extent to which the emitted spectrum falls into a sufficiently clean experimental window. In species realisations, by contrast, dark-sector dilution renders this channel ineffective for current instruments.

\begin{figure*}
    \centering
    \vs{-10mm}
    \includegraphics[width = 1.0\linewidth]{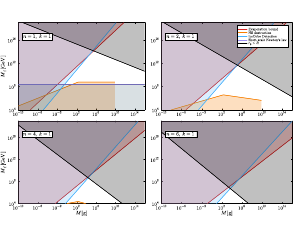}
    \vs{-15mm}
    \caption{
        The panels show the $M$--$M_{\frm}$ plane for different values of $n$ and $k = 1$ for the {\it extra-dimensional case}. The orange shaded region is excluded by neutron star existence, the red region is the evaporation bound, the ice-blue-shaded shaded region is the potential bound set by {\it IceCube} assuming a sensitivity that requires a signal of one event per year since operation start. The black shaded region shows the area of the parameter space where the size of the black holes exceeds the compactification radius $R$ and is therefore no $\mu$BH anymore. 
    }
    \label{fig:1}
\end{figure*}

\begin{figure*}
    \centering
    \vs{-10mm}
    \includegraphics[width = 1.0\linewidth]{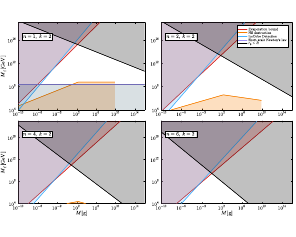}
    \vs{-15mm}
    \caption{
        Same as Fig.~\ref{fig:1}, but for $k = 2$.
    }
    \label{fig:2}
\end{figure*}

\begin{figure*}
    \vs{-10mm}
    \includegraphics[width = 1.0\linewidth]{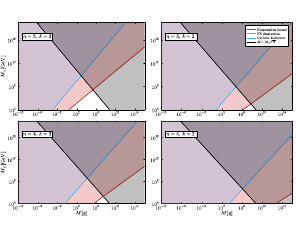}
    \vs{-15mm}
    \caption{
        The panels show the $M$--$M_{\frm}$ plane for different values of $n$ and $k$ for the {\it species case}. The colour coding is as in Figs.~\ref{fig:1} and~\ref{fig:2}. Here, the black shaded region shows the area of the parameter space where the size of the black holes exceeds $\sqrt{N}\.M_{\Prm}$ and is therefore no $\mu$BH anymore.
    }
    \label{fig:3}
\end{figure*}

\section{Neutron Star Consumption by Micro Black Holes}
\label{sec:Neutron star-Consumption-by-Micro-black holes}

\noindent To derive the first robust bounds on the model, we consider the astrophysical systems that are expected to be most sensitive to the presence of micro black hole dark matter. These are neutron stars (NSs), since they are the densest known long-lived objects that are not themselves black holes. Because of their large density, $\rho \sim 10^{-3}\,{\rm GeV}^{4}$\., and their observed lifetimes, which can reach at least $\tau_{\rm lifetime} \sim 1\,{\rm Gyr}$\., one must require that captured micro black holes do not consume a neutron star on a timescale shorter than its lifetime.

Throughout this analysis we adopt a deliberately conservative strategy. Our aim is not to derive the strongest possible constraints, but rather the most robust ones. Accordingly, whenever astrophysical or microphysical uncertainties enter, we choose assumptions that weaken the bound. The resulting exclusion regions should therefore be interpreted as conservative lower limits on the true constraining power of neutron star survival.

To estimate the effect, we follow the logic of Ref.~\cite{Giddings:2008gr}, where hypothetical stable black holes in theories with large extra dimensions were studied in the context of compact-star destruction. In the present case, however, the black holes are not strictly stable: they evaporate, but with a rate suppressed by memory burden. This modifies the balance between capture, accretion, and evaporation.

Two questions are central. First, how many $\mu$BHs are captured by a neutron star during its formation? Second, once captured, how long does it take them to consume the star? Notice that it is sufficient to consider the initial condition of the starting point
of the accretion, because neutron star consumption is a process that grows with the
mass of the accreting black holes, while the inflow of new black holes stays constant.
Even though $\mu$BHs will also be stopped by the NS after its formation, their
contribution is subleading, as they will always be in an earlier accretion phase than
the first $\mu$BHs that entered, as long as their growth rate is much larger than the
mass inflow of newly captured $\mu$BHs. This point will be addressed in more detail
below.

To answer the first question, we consider the dark matter flux during core collapse, which takes place on a timescale of order one second~\cite{Jerkstrand:2025bea}. Now it is crucial to understand what the condition is such that the $\mu$BHs
falling onto the proto-NS get captured by it and start accreting. To derive this
condition, we are going to use energy and momentum conservation. We work in units
$c=1$ and evaluate all quantities in the frame of a static observer at the
surface of the proto-NS, for which the escape velocity is
$v_\textrm{esc}=\sqrt{r_s^\textrm{NS}/R_\textrm{NS}}\simeq 0.6$, with the
associated Lorentz factor $\gamma_\textrm{esc}=(1-v_\textrm{esc}^2)^{-1/2}$.
Imagine a $\mu$BH gets dragged to the proto-NS and starts falling onto it. The Lorentz factor at the surface is
$\gamma_\textrm{in}=\gamma_\textrm{esc}\gamma_\textrm{DM}$. The corresponding
velocity when it hits the target is
\begin{equation}
    v_{\textrm{in}} = \sqrt{v_\textrm{esc}^2 +
    \gamma_\textrm{esc}^{-2}\,v_{\textrm{DM}}^2} \;,
\end{equation}
which reduces to the naive quadrature sum in the non-relativistic limit. So in
case the velocity shift is as large such that the velocity drops under
$v_\textrm{esc}$ the $\mu$BH gets trapped to the NS, this being equivalent to
the orbit becoming bound. The required shift can be calculated as
\begin{equation}
    \Delta v = \sqrt{v_\textrm{esc}^2 + \gamma_\textrm{esc}^{-2}v_{\textrm{DM}}^2}
    - v_\textrm{esc} \approx \frac{v_{\textrm{DM}}^2}{2\gamma_\textrm{esc}^2
    v_\textrm{esc}} \;.
\end{equation}
To translate this velocity shift into the mass shift, $\delta m$, the $\mu$BH has
to experience due to accretion, we impose energy and momentum conservation for
the absorption of matter at rest in the local frame,
$\gamma_\textrm{in}M+\delta m=\gamma' M'$ and
$\gamma_\textrm{in}M v_\textrm{in}=\gamma' M' v'$, which gives
\begin{equation}
    \frac{\gamma_\textrm{in}M}{\gamma_\textrm{in}M+\delta m}
    = \frac{v_\textrm{in}-\Delta v}{v_\textrm{in}}
    \;\rightarrow\;
    \frac{\delta m}{M} \approx
    \frac{v_{\textrm{DM}}^2}{2\,\gamma_\textrm{esc}\,v_\textrm{esc}^2}
    \approx 10^{-6}\;,
\end{equation}
where the last step uses $v_\textrm{DM}\simeq 10^{-3}$.
Taking this result we can estimate the mass gain of a $\mu$BH going through the forming NS by 
\begin{equation}
    \delta M = \pi r_s^2 R_\textrm{NS} \rho \;,
\end{equation}
and this mass shift has to be larger than the one derived above. Therefore the final condition we find such that a $\mu$BH gets trapped into the NS is 
\begin{equation}
    M_\textrm{f} \leq M^\frac{1-n}{2n+4} \left(\frac{M}{\delta M} R_\textrm{NS} \rho\right)^\frac{n+1}{2n+4}\;.
\end{equation}
This immediately shows that ordinary stars and planets are too dilute to capture
such objects efficiently. During core collapse, however, the density rises to values
of order $10^{12}\,{\rm g\,cm}^{-3}$, which is sufficient to reduce the stopping
length below the size of the collapsing core. As a result, the dark matter wind can
deposit $\mu$BHs into the forming proto-neutron star, where they become
gravitationally bound. Within roughly one second, the remnant reaches densities of
order $10^{14}\,{\rm g\,cm}^{-3}$, and the captured black holes then enter the
accretion regime appropriate for neutron star matter
\cite{Janka:2006fh, janka2012corecollapsesupernovaereflectionsdirections}.

Once trapped inside the relevant accretion timescale for a single black hole was identified in Ref.~\cite{Giddings:2008gr} as
\begin{align}
\begin{split}
    t
        & =  
            d_{0}\,c_{\srm}\,
            \frac{ 16\mspace{1mu}\pi\mspace{1mu}
            \bigl[
                4\mspace{1mu}\pi ( n + 1 )\.k_{n}
            \bigr]^{1/n} }{ \lambda_{n} }
            \\[2mm]
        &\phantom{ = \;}
            \times
            \left(
                \frac{ M_{\frm} }{ 1\,{\rm TeV} }
            \right)^{\!( n + 2 )/n}
            \left(
                \frac{ M_{\Prm} }{ 1\,{\rm TeV} }
            \right)^{\!2( n - 1 )/n}
            ,
\end{split}
\label{timescale}
\end{align}
with $c_{\srm}$ being the sound speed inside the neutron star, $3 < \lambda_{n} < 6.6$ for $n = 1,\dots,7$, $k_{n} \equiv 2\.(2\mspace{1mu}\pi)^{n} / [ ( n + 2 )\,\Omega_{n + 2} ]$, where $\Omega_{n + 2} \equiv 2\mspace{1mu}\pi^{n + 3/2} / \Gamma\big( n + 3 / 2 \big)$ and $d_0 = (1 \textrm{ TeV})^3/ (\pi \rho)$. Notice that during accretion within the NS, the $\mu$BH goes through various regimes, each with different timescales. The three relevant regimes are the
non-Einsteinian regime, the transition regime, and the Einsteinian regime, the
latter applying when the $\mu$BH has already grown beyond the mass given in
\eqref{eq:Micro-Black-Hole-Regime}. In \cite{Giddings:2008gr}, all formulas for
the different timescales are given, but the key finding is that the transition
regime --- under the most conservative assumption of four-dimensional accretion
behavior --- is the longest and therefore the most relevant for our considerations.
This also justifies neglecting quantum effects for $\mu$BHs that are much smaller
than nucleons, as this timescale can be calculated to be of order seconds, which
is negligible compared to the overall lifetime of $1\,\textrm{Gyr}$ for a NS.

Therefore, for the scenario to remain viable, the timescale \eqref{timescale}
must satisfy
\begin{equation}
    t
        >
            \tau_{\rm lifetime}
            \, .
\end{equation}

This criterion applies to a single captured black hole. In practice, however, a newly formed neutron star may capture multiple $\mu$BHs during the core-collapse phase. The corresponding initial number can be estimated from the incident dark matter flux as
\begin{equation}
    N_{\rm ini}
         = 
            n_{\rm DM}\,A_{\rm NS}\,v_{\rm DM} t_{\textrm{core-collapse
            }}
            \, ,
\end{equation}
where $A_{\rm NS}$ is the neutron star cross-sectional area and $v_{\rm DM}$ is the characteristic dark matter velocity. The effective destruction time is then reduced relative to the single-black hole estimate, since several seeds can accrete simultaneously.

Relative to Ref.~\cite{Giddings:2008gr}, our setup contains an additional ingredient: Hawking evaporation, modified by the memory-burden effect. Captured $\mu$BHs therefore do not simply accrete monotonically. Instead, once inside the neutron star they accrete and evaporate at the same time. Initially the black holes are assumed to lie in the memory-burden regime, but as they grow by accretion they can re-enter the semiclassical phase. The relevant question is therefore whether accretion dominates over evaporation throughout the evolution.

At the parametric level, this balance is governed by
\begin{equation}
    \frac{ \d M }{ \d t }
        \simeq
            \pi \rho\,r_{\srm}^{2} - \alpha\,T_{\Hrm}^{2}
            \, .
    \label{balancequation}
\end{equation}
For a captured black hole to grow rather than evaporate away, the right-hand side of Eq.~\eqref{balancequation} must remain positive. This condition depends on the initial mass of the infalling object: lighter black holes accrete less efficiently and radiate more strongly, whereas heavier ones accrete more efficiently and evaporate less. In our analysis we therefore impose, conservatively, that Eq.~\eqref{balancequation} be positive over the relevant part of the evolution.

We also want to check if the condition that the growth of the $\mu$BHs outweights the capture inflow justifying our ansatz that we focus on the initially bounded ones. We therefore check if the growth rate outweights the mass inflow of the captured $\mu$BHs by analyzing this equation
\begin{equation}
    \frac{dM}{dt}  = \pi \rho_{NS} r_s^2 + \rho_{DM} A_{\textrm{NS}} v_{\textrm{DM}} \;.
    \label{growthvssource}
\end{equation}
We can rewrite this equation into 
\begin{equation}
    \frac{dM}{dt} = c M^\frac{2}{n+1} + S \;,
\end{equation}
where the source term $S$ is the second term in \eqref{growthvssource} and $c = \pi \rho_{NS} M_\textrm{f}^{-2 - 2/(n+1)}$. Solving this differential equation and evaluating the deviation from the solution without the source term we get for $n = 1$ 
\begin{equation}
    t \approx t_{\textrm{no source}} - \frac{1}{c} \ln{(1+ \frac{M_\star}{M_{\textrm{0}}})} \;,
\end{equation}
with $M_\star = S/c$ for $n=1$. For $n=2$ this gets 
\begin{equation}
    t \approx t_{\textrm{no source}} - \frac{3S}{c^2 M_0^{1/3}} \;,
\end{equation}
where $M_0$ is the mass of the initially accreting black hole. Here we assumed conservatively that just one initial accreting $\mu$BH exists, and that the final mass of the NS is much higher than $M_0$. For the derived bounds it is always checked that the correction coming from the source term is never dominating. 

The same mechanism also suggests a possible connection to the missing-pulsar problem in the Galactic center. Observationally, the apparent paucity of pulsars in the central region of the Milky Way remains under debate, and it is not yet clear whether standard astrophysical effects suffice to explain it or whether additional physics is required~\cite{Pfahl:2003tf, 2010ApJ...715..939M, Dexter:2013xga}. Of course do astrophysical explanation remain viable candidates to solve the missing pulsar problem but our proposal just offers an alternative way. In the present framework, neutron stars in the Galactic center would be exposed to a substantially enhanced dark matter density, and therefore to a larger capture rate of $\mu$BHs. This can lead to efficient destruction of neutron stars in the central region while leaving those in less dense environments unaffected.

In Fig.~\ref{fig:4} we show the parameter region selected by simultaneously requiring the survival of neutron stars in the Milky Way at large and allowing efficient neutron star destruction in the Galactic center. A relatively narrow mass window is favored in this case. The underlying reason is simple: away from the Galactic center, the ambient dark matter density is too small for neutron stars to capture a sufficient number of destructive $\mu$BHs unless the individual black holes are very light. In the Galactic center, by contrast, the enhanced density allows capture and subsequent destruction even for somewhat heavier objects. Requiring both conditions simultaneously therefore isolates a restricted range of masses, centered roughly around $10^{10}\,\grm$\,.

The preferred scale of $M_{\frm}$ depends on the value of $n$. In the extra-dimensional case, values as high as
\begin{align}
    M_{\frm}
        \sim
            10^{10}\,\text{--}\,10^{11}\,{\rm GeV}
\end{align}
can remain viable for $n = 1$, as suggested for instance in dark-dimension scenarios~\cite{Montero:2022prj}. For larger $n$, the upper bound on $M_{\frm}$ becomes smaller, although lower values remain allowed as long as the micro-black hole regime is maintained. 

In order to perform this analysis, we use the Navarro-Frenk-White profile
$\rho_{\textrm{DM, gal}}(r) = \rho_s (r_s/r)\left(1+ \frac{r}{r_s}\right)^{-2}$,
with $r_s = 15.6\,\textrm{kpc}$ and $\rho_s$ fixed such that the DM density at
the solar circle is of order $0.45\,\textrm{GeV\,cm}^{-3}$. We identify the
distance from the galactic center within which pulsars get destroyed as
$1\,\textrm{kpc}$. The open parameter space shown in Fig.~\ref{fig:4} is therefore
the region where pulsars within $1\,\textrm{kpc}$ are destroyed and pulsars further
away survive. Observationally, roughly $10\%$ of the predicted pulsars within the
galactic center are still observed, so not all are destroyed completely. This does
not necessarily pose an obstacle for our scenario, as many astrophysical processes
--- such as migration or late birth of pulsars --- can account for the remaining
$10\%$. Alternatively, one could consider other profiles. An overview of currently
discussed candidates is given in \cite{Cirelli:2010xx}; the most deviating proposal
is the Burkert profile \cite{Burkert:1995yz}, which predicts a smaller overdensity
at $1\,\textrm{kpc}$ by a factor of order $6$. Smaller densities have the effect
of tightening the window, as the green band shifts to the left, indicating that the
difference in expected lifetime depending on where the pulsar resides is not large
enough to explain the deficit in the galactic center. Nevertheless, even with the
Burkert profile a small window survives in the same scenarios as in the NFW case,
since the window only starts closing for an overdensity factor in the center of
order $\sim 10$, larger than the $\sim 6$ predicted by the Burkert profile. We
therefore conclude that the micro black hole DM scenario has the potential to
address the missing pulsar problem, but a rigorous answer would require numerical
galactic simulations of NS production and DM halo development, which go beyond the
scope of this paper.

We conclude this section by noticing, that neutron star survival provides the strongest and most robust constraint considered in this work. Even under conservative assumptions, the existence of old neutron stars excludes substantial regions of parameter space. At the same time, the enhanced capture rate in the Galactic center leaves open a narrow window in which $\mu$BH dark matter can potentially contribute to the missing-pulsar problem.

\begin{figure*}
    \centering
    \vs{-10mm}
    \includegraphics[width = 1.0\linewidth]{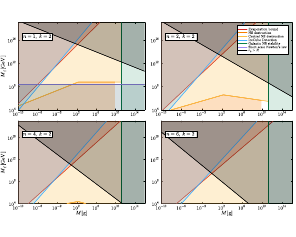}
    \vs{-15mm}
    \caption{
        The panels show the $M$--$M_{\frm}$ plane for different values of $n$ and $k = 2$ for the {\it extra-dimensional case} if one aims for a solution for the missing-pulsar problem. The orange shaded region is excluded by neutron star existence, meanwhile the yellow region is excluded as it would allow neutron stars being stable in the center of the Milky Way. The red region is the evaporation bound, the ice-blue-shaded region is the potential bound set by {\it IceCube} assuming a sensitivity that requires a signal of one event per year since operation start. The black shaded region shows the area of the parameter space where the size of the black holes exceeds the compactification radius $R$ and is therefore no $\mu$BH anymore. The green shaded region is where neutron stars are stable.
        \vs{2mm}}
    \label{fig:4}
\end{figure*}

\section{Other Gravitational Effects}
\label{sec:Other-Gravitational-Effects}

\noindent Besides the potentially dramatic consequences discussed above, $\mu$BH dark matter can also lead to additional signatures that rely purely on its gravitational interactions. Although these effects appear less constraining than neutron star survival and less promising than evaporation-based searches, they provide useful complementary diagnostics of the scenario.

A particularly interesting possibility is offered by proposed direct gravitational detectors for dark matter~\cite{Carney:2019pza, Windchime:2022whs}. Such experiments aim to detect the gravitational perturbation produced by passing compact dark matter objects on laboratory-scale sensors. This is especially relevant for the present setup, since the allowed $\mu$BH masses can lie in the mass range to which these detectors are designed to be sensitive. In contrast to short-distance tests of Newtonian gravity, such searches would probe the compact dark matter object directly and therefore constrain the $\mu$BH mass independently of the precise value of the fundamental scale $M_{\frm}$. In combination with experiments sensitive to evaporation products, this could in principle allow one to probe both the gravitational and non-gravitational manifestations of the scenario across a broad region of parameter space.

Another conceivable signature is a stochastic background of gravitational radiation from close hyperbolic encounters of $\mu$BHs. Since these objects are extremely abundant if they constitute the dark matter, one may expect a large rate of close encounters. Moreover, if the encounter takes place in the transition regime~\eqref{eq:lengthscalerange}, the effective gravitational interaction is stronger than in the ordinary Einsteinian case, so that the acceleration at closest approach is enhanced. This in turn shifts the characteristic frequency of the emitted gravitational radiation to higher values.

To estimate the size of this effect, we compare the characteristic frequency produced in a $\mu$BH encounter to the corresponding Einsteinian estimate. For an ordinary black hole, one has parametrically
\begin{equation}
    f_{\Erm}
        \sim
            \sqrt{\frac{ GM( 1 + e ) }{ r_{\Prm}^{3} }}
            \, ,
\end{equation}
where $r_{\Prm}$ is the periastron distance of the encounter and $e$ is its eccentricity. In the species case, the interaction remains effectively four-dimensional, but within the transition regime the relevant gravitational scale is set by $M_{\frm}$ rather than $M_{\Prm}$. One therefore finds the enhancement
\begin{subequations}
\begin{equation}
    \frac{ f_{\mu{\rm BH}}^{\rm species} }{ f_{\Erm} }
        \sim
            \frac{ M_{\Prm} }{ M_{\frm} }
            \, .
\end{equation}
In the extra-dimensional case, dimensional analysis similarly gives
\begin{equation}
    \frac{ f_{\mu{\rm BH}}^{\rm extra} }{ f_{\Erm} }
        \sim
            \left(
                2\mspace{1mu}\pi\,\frac{ R }{ r_{\Prm} }
            \right)^{\!n/2}
            .
\end{equation}
\end{subequations}

In both realisations the characteristic frequencies are therefore significantly enhanced relative to the already extremely large Einsteinian values. For example, an ordinary black hole of mass $10^{8}\,{\rm g}$ would already produce frequencies of order $\Ocal( 10^{26} )\,{\rm Hz}$ in a close encounter, even before including the short-distance enhancement. Current proposals for high-frequency gravitational-wave detectors aim at sensitivities no higher than roughly $10^{18}\,{\rm Hz}$~\cite{Aggarwal:2020olq}, and even these projections remain under active discussion~\cite{TitoDAgnolo:2024res}. It therefore appears very unlikely that gravitational radiation from $\mu$BH encounters can be observed with present or near-future technology.

In summary, purely gravitational probes of $\mu$BH dark matter are conceptually interesting and potentially complementary to evaporation-based searches. Direct gravitational detection of passing compact dark matter may eventually become relevant, whereas gravitational-wave signatures from close encounters are pushed to frequencies far beyond realistic experimental reach. Within the class of observables considered here, these effects therefore do not provide the leading constraints or discovery prospects.

\section{Micro Black Holes Mergers}
\label{sec:Micro-black holes-Mergers}

\noindent In a universe populated by extremely light micro black holes, mergers can occur much more frequently than in conventional primordial-black hole scenarios because of the enormous number density of the objects. In the present framework, such events are of particular interest because $\mu$BHs can act as dark matter only if their evaporation is sufficiently suppressed by the memory-burden effect. If, however, the merger of two memory-burdened black holes resets the remnant to a semiclassical state, as suggested in Refs.~\cite{Zantedeschi:2024ram, Dondarini:2025ktz}, then the newly formed black hole can briefly evaporate much more rapidly before re-entering the memory-burden regime.

This possibility is phenomenologically important. Since the merging black holes are extremely light, the post-merger remnant is correspondingly hot and can emit highly energetic particles. In that case, the very mechanism that renders the progenitor black holes cosmologically stable temporarily ceases to operate, and the merger event becomes a source of visible evaporation products. The resulting signal depends on how long the remnant remains in the semiclassical phase and what fraction of its mass is emitted before memory burden is re-established.

For a monochromatic primordial-black hole mass function in which the binaries decouple before matter-radiation equality, the merger rate has been estimated conservatively as~\cite{Inman:2019wvr, Hutsi:2020sol, Franciolini:2022htd}
\begin{equation}
    R_{\rm PBH}
         \sim 
            f_{\rm PBH}^{53/37}
            \left(
                \frac{ t_{0} }{ t }
            \right)^{\!34/37}\mspace{-2mu}
            \left(
                \frac{ 2\,m_{\rm PBH} }{ 10^{10}\,\grm }
            \right)^{\!-32/37}
            \frac{ 10^{-68} }{ {\rm cm}^{3}\,\srm }
            \, .
\end{equation}
The quantity of direct observational interest is the flux at Earth of particles emitted by these merger remnants. Following Refs.~\cite{Zantedeschi:2024ram, Dondarini:2025ktz}, the Galactic contribution can be written as
\begin{equation}
    \frac{ \d\Phi_{i} }{ \d E_{\rm Gal} }
        \simeq
            q\.\tau_{\rm sc}
            \int \frac{ \d\Omega }{ 4\mspace{1mu}\pi }
            \int \d s\;R_{\rm PBH}\,
            \delta\bigl[ r( s,\mspace{1.5mu}\theta ) \bigr]\,
            \frac{ \d^{2} N_{i}( E ) }{ \d E\,\d t }
\end{equation}
and the extragalactic contribution as
\begin{equation}
    \frac{ \d\Phi_{i} }{ \d E_{\rm eg} }
        \simeq
            \frac{ q\.\tau_{\rm sc} }{ 4\mspace{1mu}\pi }
            \int_{0}^{z_{\frm}}\d z\,
            \bigg| \frac{ \d t }{ \d z } \bigg|\;
            R_{\rm PBH}\bigl[ t( z ) \bigr]\,
            \frac{ \d^{2} N_{i}( E' ) }{ \d E\,\d t }\, ,
\end{equation}
where $\delta( r ) \equiv \rho_{\rm DM}( r )/\rho_{\rm DM}$ describes the enhancement of the dark matter density toward the Galactic center, and where a Navarro--Frenk--White profile is assumed~\cite{Pujolas:2021yaw, Navarro:1996gj}. The quantity $r( s,\mspace{1.5mu}\theta )$ denotes the line-of-sight distance of the merger from Earth, $\d t/\d z$ is the cosmological line element, and $E' = E( 1 + z )$ accounts for the redshift between emission and observation.

In the $\mu$BH scenario considered here, several aspects of this calculation differ from the ordinary four-dimensional case studied in Refs.~\cite{Zantedeschi:2024ram, Dondarini:2025ktz}. The first important distinction arises in generic species models. In order for $\mu$BHs to constitute dark matter in such realisations, the number of species must typically be very large, since the transition range~\eqref{eq:Micro-Black-Hole-Regime} must be sufficiently broad for the modified short-distance behavior to be phenomenologically relevant. However, in precisely this regime Eq.~\eqref{eq:E-Ratio} implies that only a parametrically small fraction of the evaporated energy is released into observable Standard Model particles. Most of the energy is emitted into dark sectors. As a consequence, merger-induced signals in species scenarios are expected to be far below current observational sensitivities.

The situation is more favorable in extra-dimensional realisations, where the visible-sector branching fraction is not comparably suppressed. In that case, the merger-induced flux must nevertheless be recalculated using the modified semiclassical lifetime in the transition regime. Specifically, one must replace the ordinary four-dimensional lifetime by Eq.~\eqref{eq:tau-Extra-Dimensions} and similarly use the appropriate expression for the parameter $q$, which determines the fraction of the remnant mass emitted before the memory-burden phase resumes. This distinction is not very important when $q \sim \Ocal( 1 )$, but it becomes significant if $q$ scales as an inverse power of the entropy, since the entropy itself is modified in the $\mu$BH regime according to Eq.~\eqref{eq:S-Extra-Dimensions}.

In Fig.~\ref{fig:5} we show the resulting neutrino fluxes at Earth for representative parameter choices motivated by currently discussed extra-dimensional scenarios~\cite{Arkani-Hamed:1998jmv, Montero:2022prj}. For the black hole masses we choose the lightest values allowed by the analysis of Sec.~\ref{sec:Neutron star-Consumption-by-Micro-black holes}, since these yield the hottest remnants, which are the strongest bounded ones by experiments. This choice is conservative from the perspective of detectability, because the highest-energy evaporation products are typically the easiest to distinguish from backgrounds.

In the original ADD framework, where $M_{\frm} \sim \Ocal( {\rm TeV} )$, the first two benchmark cases correspond to $n = 2$ and $n = 6$. For $n = 2$, the expected neutrino energies lie in the GeV range, where the astrophysical background is large and the observational separation between signal and background is therefore weakest. For $n = 6$, the Hawking temperature is pushed closer to the fundamental scale, and the detection prospects improve accordingly. We also consider higher-scale benchmarks with $M_{\frm} = 10^{10}\,{\rm GeV}$ and $n = 1$ or $n = 2$, as often discussed in dark-dimension scenarios~\cite{Montero:2022prj}. In these cases the expected flux for $q = \Ocal( 1 )$ approaches current bounds most closely, with the most promising benchmark occurring for $n = 2$. This suggests that merger-induced evaporation bursts could become observationally relevant in future searches for a restricted portion of the extra-dimensional parameter space.

If, on the other hand, the remnant re-enters the memory-burden regime almost immediately after the merger, then only a very small fraction of its mass is radiated in the semiclassical phase and the observable flux is strongly reduced. In that limit, black hole mergers do not provide a competitive probe of $\mu$BH dark matter.

In summary, merger-induced evaporation is a potentially important signature only if two conditions are met: the post-merger remnant must temporarily return to the semiclassical regime, and a non-negligible fraction of its mass must be emitted before memory burden turns on again. These conditions appear difficult to realise in generic species models because of dark-sector dilution, but they can be satisfied in some extra-dimensional scenarios, where the resulting high-energy flux may approach observational relevance.

\begin{figure}
    \vs{4mm}
    \includegraphics[width = 1.0\linewidth]{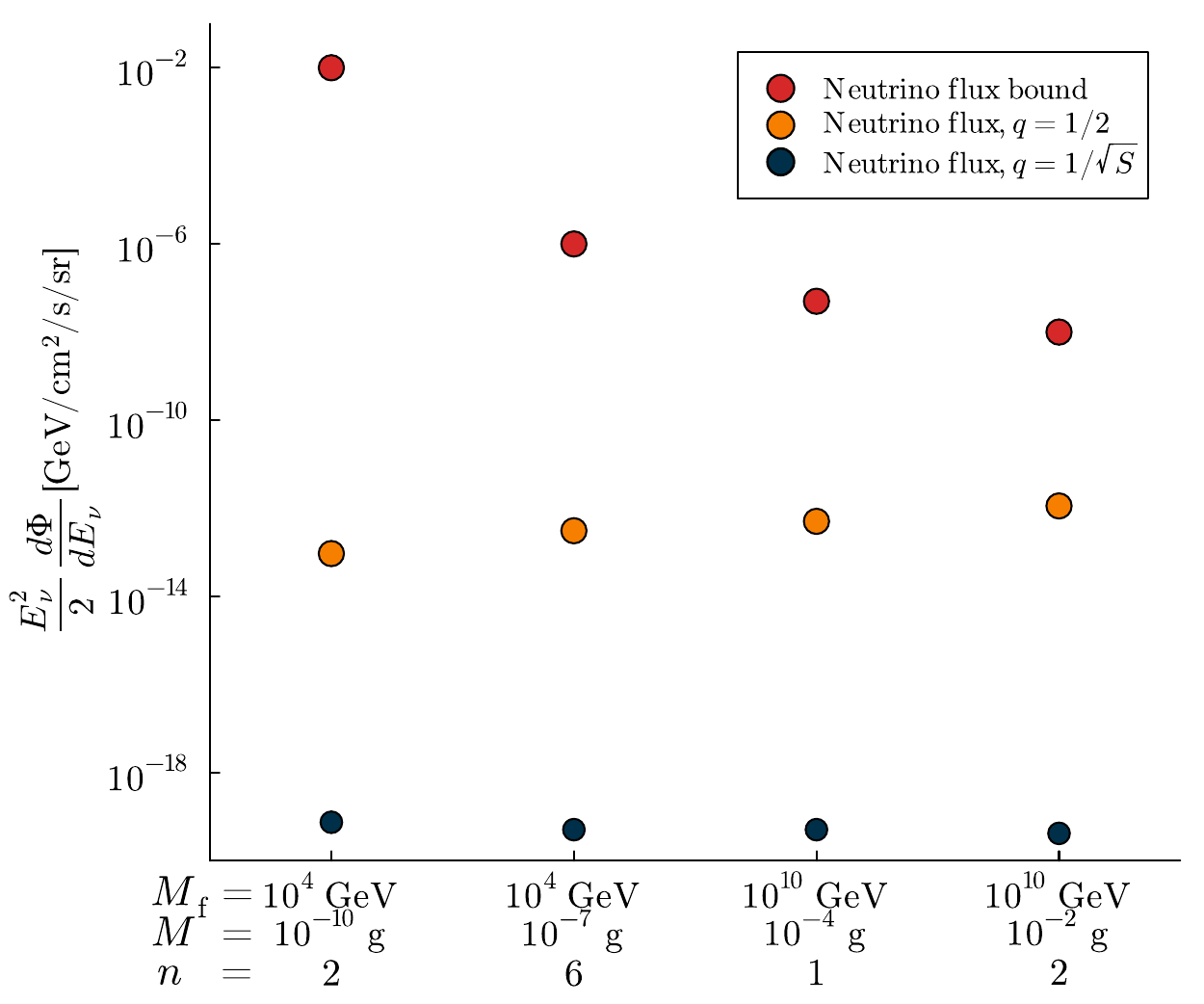}
    \caption{
        The flux of evaporated neutrinos on the Earth for different scenarios of {\it extra-dimensional theories}. Two scales for $M_{\frm}$ are investigated, motivated by Refs.~\cite{Arkani-Hamed:1998jmv, Montero:2022prj}. The mass of the black holes has been chosen to be the lightest value allowed for such a scenario as they lead to the highest flux. The red dots show the bound on neutrinos which have the energies of the evaporated neutrinos. The orange dots show the flux in case half of the black hole evaporates before it reenters the memory burden phase. The blue dots show the same in case just a fraction of $1/\sqrt{S}$ evaporates. The bounds of the fluxes are stemming from Refs.~\cite{Super-Kamiokande:2015qek, ANTARES:2024ihw}.
    }
    \label{fig:5}
\end{figure}

\section{Conclusion}
\label{sec:Conclusion}

\noindent We have studied the phenomenology of primordial micro black holes as dark matter in frameworks where gravity is modified at short distances by either extra dimensions or a large number of species, under the assumption that the memory-burden effect suppresses Hawking evaporation strongly enough to stabilise even very light black holes on cosmological timescales. This setup opens a qualitatively new dark matter regime. Once black holes in the transition region between the fundamental scale $M_{\frm}$ and the Einsteinian regime survive until today, their tiny masses imply enormous number densities and therefore potentially observable interaction rates with astrophysical environments and terrestrial detectors.

Our analysis shows that neutron star survival provides the strongest and most robust probe of the scenario. Even under conservative assumptions, the existence of long-lived neutron stars excludes substantial regions of parameter space. At the same time, the enhanced capture rate expected in the Galactic center leaves open a narrow region in which $\mu$BH dark matter can potentially contribute to the missing-pulsar problem. In this sense, neutron stars are not only a constraint, but also a potentially informative probe of the underlying dynamics.

The direct evaporation signal from the ambient $\mu$BH population is much more model-dependent. In extra-dimensional realisations, the combination of large number density and sufficiently energetic Hawking products can make neutrino telescopes sensitive to the lightest viable black holes, especially when the visible-sector emission is not strongly diluted and the characteristic energies fall into a relatively clean observational window. In generic species scenarios, by contrast, evaporation into Standard Model particles is strongly suppressed relative to the total emitted power, making diffuse evaporation searches ineffective with present-day instruments.

We have also considered merger-induced signatures. If the merger of two memory-burdened black holes resets the remnant to a semiclassical evaporation phase, then a population of very light $\mu$BHs can generate bursts of energetic particles. This channel is again most promising in extra-dimensional models, particularly when the remnant loses an $\Ocal( 1 )$ fraction of its mass before re-entering the memory-burden regime. If instead the remnant returns almost immediately to the memory-burden phase, the observable flux is strongly reduced and current constraints are easily avoided. Purely gravitational probes, such as direct detection of passing compact dark matter or gravitational radiation from hyperbolic encounters, are interesting conceptually but appear less promising with present or near-future sensitivities.

Overall, our results indicate that micro black hole dark matter is not automatically excluded once memory burden and modified short-distance gravity are taken seriously. Rather, it defines a constrained but nontrivial phenomenological window, with markedly different observational prospects in extra-dimensional and species realisations. Neutron star survival currently provides the most powerful handle on the scenario, while neutrino telescopes and merger-induced evaporation products offer complementary avenues for testing it. A more refined treatment of detector response, compact-object capture, merger histories, and post-merger memory-burden dynamics should further sharpen the viability of this class of models.

\noindent
{\it Acknowledgments\,---} M.E.~acknowledges stimulating discussions with Hyungjin Kim and Nikola Savi{\'c}. We are also thankful for the helpful comments of the anonymous referee. The work of M.E.~was supported by ANR grant ANR-23-CE31-0024 EUHiggs.

\bibliography{refs}

\if{case}
\begin{figure*}
    \centering
    \vs{-10mm}
    \includegraphics[width = 1.0\linewidth]{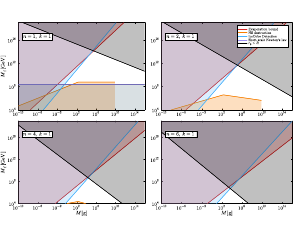}
    \vs{-15mm}
    \caption{
        The panels show the $M$--$M_{\frm}$ plane for different values of $n$ and $k = 1$ for the extra-dimensional case. The orange shaded region is excluded by neutron star existence, the red region is the evaporation bound, the ice-blue-shaded shaded region is the potential bound set by {\it IceCube} assuming a sensitivity that requires a signal of one event per year since operation start. The black shaded region shows the area of the parameter space where the size of the black holes exceeds the compactification radius $R$ and is therefore no $\mu$BH anymore. 
    }
    \label{fig:1}
\end{figure*}

\begin{figure*}
    \centering
    \vs{-10mm}
    \includegraphics[width = 1.0\linewidth]{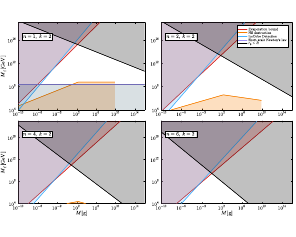}
    \vs{-15mm}
    \caption{
        The panels show the $M$--$M_{\frm}$ plane for different values of $n$ and $k = 2$ for the {\it extra-dimensional case}. The orange shaded region is excluded by the existence of neutron star, the red region is the evaporation bound, the ice-blue-shaded region is the potential bound set by {\it IceCube} assuming a sensitivity that requires a signal of one event per year since operation start. The black shaded region shows the area of the parameter space where the size of the black holes exceeds the compactification radius $R$ and is therefore no $\mu$BH anymore.
    }
    \label{fig:2}
\end{figure*}

\begin{figure*}
    \centering
    \vs{-10mm}
    \includegraphics[width = 1.0\linewidth]{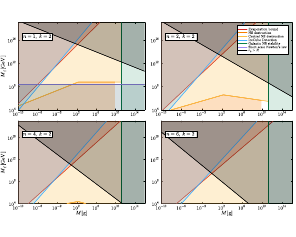}
    \vs{-15mm}
    \caption{
        The panels show the $M$--$M_{\frm}$ plane for different values of $n$ and $k = 2$ for the {\it extra-dimensional case} in case one aims for a solution for the missing-pulsar problem. The orange shaded region is excluded by neutron star existence, meanwhile the yellow region is excluded as it would allow neutron stars being stable in the center of the Milky Way. The red region is the evaporation bound, the ice-blue-shaded region is the potential bound set by {\it IceCube} assuming a sensitivity that requires a signal of one event per year since operation start. The black shaded region shows the area of the parameter space where the size of the black holes exceeds the compactification radius $R$ and is therefore no $\mu$BH anymore. The green shaded region is where neutron stars are stable.}
    \label{fig:3}
\end{figure*}

\begin{figure}
    \vs{-10mm}
    \includegraphics[width = 1.0\linewidth]{Figures/MergerFlux.pdf}
    \caption{
        The flux of evaporated neutrinos on the Earth for different scenarios of {\it extra-dimensional theories}. Two scales for $M_{\frm}$ are investigated, motivated by Refs.~\cite{..., ...}. The mass of the black holes has been chosen to be the lightest value allowed for such a scenario as they lead to the highest flux. The red dots show the bound on neutrinos which have the energies of the evaporated neutrinos. The orange dots show the flux in case half of the black hole evaporates before it reenters the memory burden phase. The blue dots show the same in case just a fraction of $1/\sqrt{S}$ evaporates.
    }
    \label{fig:4}
\end{figure}

\begin{figure*}
    \vs{-10mm}
    \includegraphics[width = 1.0\linewidth]{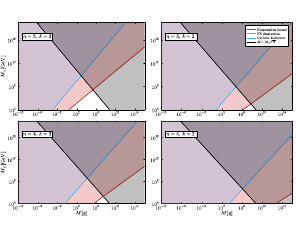}
    \vs{-15mm}
    \caption{
        The panels show the $M$ and $M_{\frm}$ plane for different values of $n$ and $k$ for the {\it species case}. The orange shaded region is excluded by neutron star existence, the red region is the evaporation bound, the ice-blue-shaded shaded region is the potential bound set by {\it IceCube} assuming a sensitivity that requires a signal of one event per year since operation start. The black shaded region shows the area of the parameter space where the size of the black holes exceeds $\sqrt{N}\.M_{\Prm}$ and is therefore no $\mu$BH anymore.
    }
    \label{fig:5}
\end{figure*}
\fi

\end{document}